\documentclass[10pt,letterpaper,twocolumn]{article} %% two column, final layout

\usepackage{ol2}
\usepackage[draft]{hyperref}
\usepackage{amsmath}

\begin{document}

\twocolumn[ %% activate for two-column option

\title{A bias free true random number generator}

%% For REVTeX it is possible to automate superscript and e-mail callouts with the superscriptaddress option; see REVTeX4 documentation.

\author{Wei Wei and Hong Guo$^*$}

\address{
CREAM Group, State Key Laboratory of Advanced Optical Communication
Systems and Networks and Institute of Quantum Electronics, School of
Electronics Engineering and Computer Science, Peking University,
Beijing 100871, P.R. China\\
$^*$Corresponding author: hongguo@pku.edu.cn}

\begin{abstract}We propose a new approach to nondeterministic random number
generation. The randomness originated from the uncorrelated nature
of consecutive laser pulses with Poissonian photon statistics and
that of photon number detections is used to generate random bit, and
the von Neumann correction method is used to extract the final
random bit. This method is proved to be bias free in randomness
generation provided that the single photon detections are mutually
independent. Further, it has the advantage in fast random bit
generation since no postprocessing is needed. A true random number
generator based on this method is realized and its randomness is
tested and guaranteed using three statistical test batteries.
\end{abstract}


 ] %% activate for two-column option

\noindent Random numbers are essential in an extremely wide
application range, such as statistical sampling \cite{Lohr},
computer simulations \cite{Gentle}, randomized algorithm
\cite{Mitzenmacher} and cryptography \cite{Menezes+}. Pseudo random
numbers generated with certain algorithms are widely used in modern
digital electronic information systems. Though the pseudo random
number generators (PRNG), thanks to the development of computer
science and technology, have been highly refined in terms of
generation rate and robustness against random test, the algorithm
based PRNG can not generate true randomness and so has the essential
drawback in the applications which require physical randomness, such
as quantum cryptography, which requires unpredictable bit strings to
ensure the inability of estimation \cite{Gisin+}. True random
numbers should be unpredictable and unreproducible. For this reason,
physical random phenomena, such as the decay of radioactive
nucleus\cite{walker}, thermal noise in resistors \cite{Holman},
frequency jitter of electronic oscillators \cite{Bucci} and photon
emission noise \cite{Stipcevi+,toshiba}, etc., are used as physical
sources for nondeterministic random number generations. Random
number generator based on these physical random processes are termed
as true random number generators (TRNG), and those based on photon
emission noise extract random bit either from the random time
intervals between photon emissions of semiconductors
\cite{Stipcevi+}, or from the collapse of the photon wave function
on random gating cycle \cite{toshiba}. In these schemes, the timing
precision is a limitation to the random bit generation rate.
\begin{figure}[htb]
\centerline{\includegraphics[width=8cm]{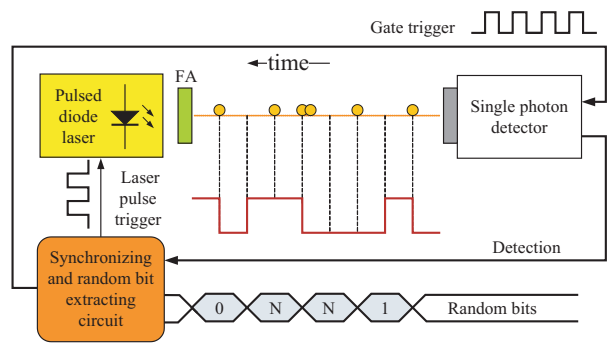}} \caption{(color
online) Schematic experiment set up of our TRNG. FA: flexible
attenuator. 1: bit \textbf{1}. 0: bit \textbf{0}. N: not used.}
\end{figure}

In this Letter, we propose a new approach to realize a TRNG based on
photon number detection of weak laser pulse. The main advantage of
this new type of TRNG is that it has equal probabilities for the
bits of ones and zeros and is bias free from the variability of the
device and the environment. Moreover, this new type of TRNG can be
realized with a simple and compact setup. It can generate random
numbers at a high speed and is only limited by the repetition rate
of the single photon detector (SPD). These advantages make it
applicable for quantum cryptography and other applications which
require fast generation of true random bits.

According to quantum theory of laser, the photon statistics for
laser operating above threshold approaches to
Poissonian\cite{walls}, indicating that laser light has the unity
second order correlation function \cite{walls} and the photon
numbers of different laser pulses are mutually independent. A TRNG
can be implemented based on this fact. Since the photon number
distribution of partially absorbed light follows a \emph{Bernoulli
transform} of the initial field \cite{Leonhardt}, provided that the
detections are mutually independent, it can be proved that the
photon numbers detected by a photodetector from weak laser pulses
are also Poissonian distributed, i.e., $P_\eta  (n) = (\eta
\lambda)^n \exp(-\eta \lambda)/n!$, where ${\lambda }$ is the mean
photon number of the weak laser pulses, and ${\eta }$ is the
detection efficiency of the photodetector. The probabilities of
$n_1>n_2$ and $n_1<n_2$ are equal, where $n_1$ and $n_2$ are the
photon numbers of two consecutive pulses. In experiment, we use an
avalanche photodiode (APD) operating in gated Geiger mode for the
photon detection, which does not distinguish the photon numbers
above zero. Thus, we code ``$n_1 > 0$ and $n_2 = 0$" as bit
\textbf{1} and ``$n_1 = 0$ and $n_2 > 0$" as bit \textbf{0}. Then,
for two consecutive detections, the probabilities $P(\textbf{1})$
and $P(\textbf{0})$ of generating a bit \textbf{1} and a bit
\textbf{0}, respectively, follow
\begin{align}
P(\textbf{1}) =P(\textbf{0})=P_{\eta} (n>0) \times P_{\eta}
(n=0)=e^{ - \eta \lambda }(1 - e^{ - \eta \lambda }).
\end{align}
It is then evident that the probabilities of generating the bits of
ones and zeros are {\it intrinsically} equal and hence, a TRNG based
on this method is physically unbiased.

The experiment setup of our TRNG is shown in Fig. 1. Laser pulses
(300ps@1550nm) are generated by a pulsed distributed feedback (DFB)
diode laser (id300, id Quantique), and the mean photon number of
each pulse can be continuously adjusted by a flexible attenuator
(FA). Photon number is detected by an InGaAs APD based SPD (id200,
id Quantique) working in gated Geiger mode. The gate width of 2.5ns
is chosen in experiment to minimize the probability of dark count.
We develop an FPGA-based circuit for system synchronization and
random bit extraction. The true random bits are derived from the
sequence of detection signals using von Neumann correction method
\cite{vonNeumann}. Using this method, we scan the random sequence
from left to right reading successive pairs of photon detections.
The outputs of ``$n_1>0$ and $n_2=0$" and ``$n_1=0$ and $n_2>0$" are
adopted as the bits \textbf{1} and \textbf{0}, respectively, while
those of ``$n_1>0$ and $n_2>0$" and ``$n_1=0$ and $n_2=0$" are
abandoned, though they also contain randomness that can be extracted
by other correction procedures (see, e.g., \cite{Peres}). However,
for doing this, some postprocessing is needed, which may
\emph{greatly} reduce the random bit generation rate. We, instead,
adopt {\it von Neumann correction method} for our TRNG because it is
simple and easy to implement with digital logic and no
postprocessing is needed. The repetition frequency of the generation
and detection of laser pulses is chosen as 1 MHz, and the dark count
probability for our SPD is measured to be 3$\times$10${^{-5}}$ with
a gate width of 2.5ns. The detection efficiency of SPD is $\sim$10\%
(specification of id Quantique).
\begin{figure}[htb]
\centerline{\includegraphics[width=8cm]{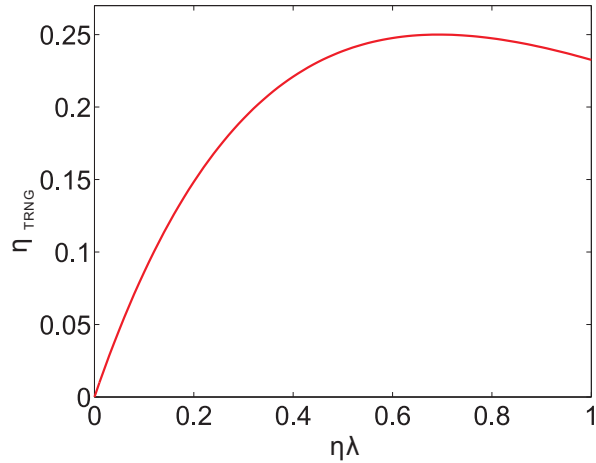}}
\caption{(color online) Random bit efficiency $\eta_{\rm TRNG}$ as a
function of $\eta\lambda$. $\eta$: detection efficiency of SPD;
$\lambda$: mean photon number of each laser pulse.}
\end{figure}

To ensure that our TRNG generates true random bits, we must confirm
that every single photon detection of the laser pulses is mutually
independent. In detection, the after-pulse effect of APD may cause
some correlation between two consecutive detection results. For an
APD operating in gated Geiger mode, carriers are trapped after every
avalanche. If some of them emit during the next gate-on time, it can
trigger a new avalanche (after-pulse) even though no photon is
absorbed. For the detection with InGaAs APDs, after-pulse is
frequently found when the repetition frequency exceeds 1 MHz. Since
only traps with an emission lifetime comparable to or longer than
the time interval between two consecutive gates generate
after-pulses, we can introduce a dead time following an avalanche,
during which, no gate is applied to the APD. This is an effective
way to eliminate the after-pulse effect \cite{Stuchi}. For the SPD
of our experiment, a dead time of 8 $\mu$s is chosen, and our
experiment results show that the probability of after-pulse is
negligible and the non-signal counts come only from the dark counts.

A technically important issue for TRNG is its efficiency to extract
true random bit from random events, $\eta_{\rm TRNG}$, which is
defined as the number of random bits per random event. For our TRNG,
a random event is a photon number detection of a weak laser pulse.
Thus, $\eta_{\rm TRNG}$ is equal to half of the probability that a
random bit is generated by a pair of detections, i.e.,
\begin{equation}
\eta_{\rm TRNG}=\frac{P(\textbf{1})+P(\textbf{0})}{2}=e^{ - \eta
\lambda }(1 - e^{ - \eta \lambda }).
\end{equation}
When $\eta\lambda={\rm ln2}\approx0.693$, we get the optimal
$\eta_{\rm TRNG}$ as 0.25. This is also the point that the
probability of detecting zero or above zero photons are equal. The
value of $\eta_{\rm TRNG}$ varying with $\eta\lambda$ is illustrated
in Fig. 2. It shows that when $\eta_{\rm TRNG}$ is at its maximum,
it is not sensitive to slight fluctuations of $\eta\lambda$ around
0.693. Experimentally, we adjust the flexible attenuator to the
state where the probabilities of detecting zero and above zero
photons are approximately equal. In this case, our TRNG generates
intrinsically unbiased random bits at the optimal rate permitted by
the current setup.

We use three batteries of statistical tests to evaluate the
performance of our TRNG. They are ENT \cite{hotbits}, DIEHARD
\cite{diehard} and STS \cite{STS}. We record a bit file of 9.45
$\times$ 10$^8$ bits for the tests. The ENT results of our TRNG are:
Entropy = 1.000000 bits per bit and the optimum compression would
reduce the bit file by 0 percent, $\chi^2$ distribution is 1.49, and
randomly would exceed this value 22.17\% of the times, arithmetic
mean value of data bits is 0.5000 (0.5 = random), Monte Carlo value
for $\pi$ is 3.141747714 (error 0.00 percent), serial correlation
coefficient is 0.000041 (totally uncorrelated = 0.0). A DIEHARD
test, with a data sample of 10 $\sim$ 11 Mbytes, is considered
succeed if $0.01\leq p \leq 0.99$ is satisfied \cite{diehard}. For
multiple $p$-values cases, we use a KS test to obtain a final
$p$-value and the test result is based on the final $p$-value. The
results of DIEHARD test for our TRNG are illustrated in Fig. 3 and
show that all final $p$-values are within 0.01$\sim$0.99. In STS
test, the large experimental file is divided into 920 separate
smaller streams of 10$^6$ bits. An individual bit stream is usually
considered to pass a particular test when $p\geq 0.01$ and
consequently, 98\% $\sim$ 100\% of all the bit streams are expected
to pass a particular test due to statistical fluctuations
\cite{STS}. Fig. 4 shows that all tests are passed with the final
$p$-values (after KS test) above the significance level, and all
testing results are within the confidence interval for the
proportion of pass. The above three random tests show that our TRNG
has good quality in randomness.
\begin{figure}[htb]
\centerline{\includegraphics[width=8cm]{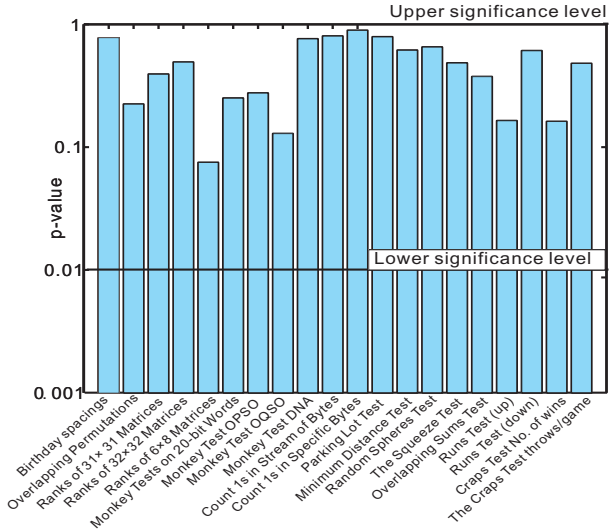}}
\caption{(color online) Results of DIEHARD. All testes are passed at
lower (upper) significance level of 0.01 (0.99).}
\end{figure}
\begin{figure}[htb]
\centerline{\includegraphics[width=8cm]{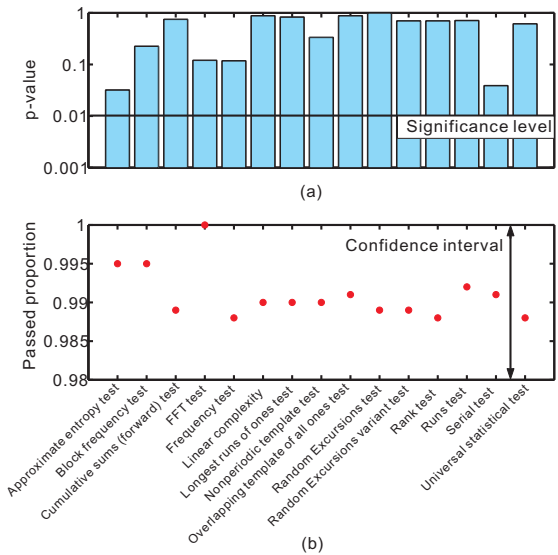}} \caption{(color
online) Results of STS. All testes are passed at the significance
level of 0.01.}
\end{figure}

We propose a new approach to realize a TRNG that is intrinsically
bias free. Its randomness is based on the photon number detection of
weak laser pulses. Compared with other bias free TRNGs, e.g., those
based on photon emission noise, the complicated timing circuits with
high precision is not needed in our TRNG and thus is convenient to
be implemented. Currently, the random bit generation rate of our
TRNG is restricted only by the repetition rate of our InGaAs APD
based SPD. Hence, a faster TRNG can be experimentally implemented
with a faster, say, silicon APD based SPD and the random bit
generation rate can be up to Gbit/s, which is suitable for high
speed applications.

%section{acknowledgement}
This work is supported by the Key Project of National Natural
Science Foundation of China (Grant No. 60837004). We are grateful to
X. Peng, W. Jiang, J. W. Zhang, T. Liu, J. Yang, Z. G. Zhang and F.
Grosshans for their helps and suggestions during drafting this
manuscript.

\end{document}